\documentclass{appolb}
\usepackage{epsfig}
\usepackage{color}
\usepackage{cite}
\begin{document}
\pagestyle{plain}
\newcount\eLiNe\eLiNe=\inputlineno\advance\eLiNe by -1
\title {Consecutive  partitions of social networks between rivaling leaders}
\author{Ma{\l}gorzata J. Krawczyk$^1$, Krzysztof Ku{\l}akowski$^1$\footnote{e-mail: kulakowski@fis.agh.edu.pl},\\ Janusz A. Ho{\l}yst$^2$ $^3$
\address{$^1$ Faculty of Physics and Applied Computer Science, AGH University of Science and Technology, al. Mickiewicza 30, 30-059 Cracow, Poland\\
$^2$ Faculty of Physics, Warsaw University of Technology, ul. Koszykowa 75, 00-662 Warsaw, Poland\\
$^3$ ITMO University, Kronverskiy av. 19, RU197101 Saint Petersburg, Russia}}

\maketitle

\maketitle

\begin{abstract}
A model algorithm is proposed to study subsequent partitions of complex  networks describing social structures. The partitions are supposed to appear as actions of rivaling leaders corresponding to nodes with large degrees. The condition of a partition is that the distance between two leaders  is at least three links. This ensures that the layer of 
nearest neighbours of each leader remains attached to him. As a rule, numerically calculated size distribution of fragments of scale-free Albert-Barabasi networks reveals one large fragment which contains the original leader (hub of the network), and a number of small fragments with opponents  that are described by two Weibull distributions. Numerical simulations and mean-field theory reveal that size of the larger fragment scales as the square root of the initial network size.  The algorithm is applied to the data on political blogs in U.S. (L. Adamic and N. Glance, Proc. WWW-2005). The obtained fragments are clearly polarized; either they belong to Democrats, or to the GOP.
\end{abstract}

\section{Introduction}

The separation of opposites is the earliest theory of creation of the World, as it is ascribed to Anaximander (c. 609 - c. 547 BC) \cite{tatar}. Appearance of opposition against current leadership was natural also for the social philosophy of Thomas Hobbes (1588-1679), best known for the motto 'homo homini lupus' \cite{tatar}. Cyclic instabilities of political power, when new leaders emerge, are known to appear throughout history \cite{heat,duin}. In modern times, the concept moved to specific branches of knowledge, from biology \cite{b1,b2} to game theory \cite{barth}; it remains vivid also in political science \cite{specu}. Stalinist strategy of successive exclusion of moderate political groups has been clearly described in \cite{kers}. In \cite{raf}, examples can be found of successive splintering of terrorist groups in Northern Ireland and Near East. While the concept of separation is valid almost everywhere, its relevance for social sciences is supported by its possible implications for numerous problems in social psychology \cite{zac,sus,opp}. \\

The problem of partition of a network into communities is well established in the socio-physical literature \cite{for,T1}. As a rule, communities are understood there as subsets of the network more densely connected than the network as a whole. Here we propose and explore a specific model of partition, motivated by selection of leaders \cite{L1,L2,L3}. Here, the function of leaders is
assigned to the nodes with the largest possible degree; this assumption is consistent with the literature \cite{wfst,ball}. An additional condition is that nearest neighbours of each leader remain attached to him during all stages of the partition. Setting this condition, we are lead by 'the 11-th Law of the Inner Cycle' by John C. Maxwell: 'a leader's potential is determined by those closest to him' \cite{jcm}, which highlights the validity of ties between a leader and his closest circle. As an additional argument, we can indicate historical examples, when a leader finds a group of supporters who remain faithful to him even after he is defeated \cite{hb,is}.\\

The network structure used here is the Albert-Barabasi one, for its associations with at least some social structures \cite{sciam}. Our aim here is to divide the network into fragments, centered around local leaders i.e. nodes with highest connections degrees. In each fragment, two main  leaders are identified, and the split is simulated again. The algorithm terminates, when each obtained fragment cannot be divided according to the above conditions. As an outcome, we get the size distribution of the obtained fragments of the network and the number of successive partitions as dependent on the network density. \\

In the next section the algorithm of the partition is explained in details. Third section is devoted to the numerical results, and fourth section to analytical estimation of    largest  component of parted network.  In Section 5, we report the application of the algorithm to the data on political blogs in U.S. \cite{adamic}. Discussion and summary are given in the last section.

\section{The algorithm}

A scale-free Albert-Barabasi network of $N$ nodes is constructed in the standard way \cite{bk}, with the number of links from a new node to old nodes equal $M=1,2,3$. Links of the network are numbered according to the order in time; those added later have larger number. For links added simultaneously, i.e. with the same node, their mutual order is not relevant. A node is found with the largest degree $k_m$, and it is marked as the first leader. The rival leader is found as the node with maximal degree, less or equal to $k_m$, such that the distance to the first leader is not smaller than three links.\\

The process of cutting links starts from a selection of the shortest path between the leaders. If there is more than one path, we concentrate on one of them. If the length of the path is exactly three, there is only one link in the middle to be cut. If the shortest path consists of more than three links, the cutting can be performed in two ways; either we select the link with the lowest number (variant A) or the link with the highest number (variant B). Comparing the results of both, we will be able to state, how the selection of the link age is relevant. The process is repeated: again and again the shortest path between the leaders is found and one of the links is cut. \\

When the network is split into two, there is one leader in each fragment. For each of them we appoint a new rival leader with the same method as above. Then, the procedure of splitting is repeated. If for all nodes, the shortest path from a node to the leader is less than three links, the algorithm is stopped.\\

\begin{figure}[!hptb]
\begin{center}
\includegraphics[width=.99\columnwidth, angle=0]{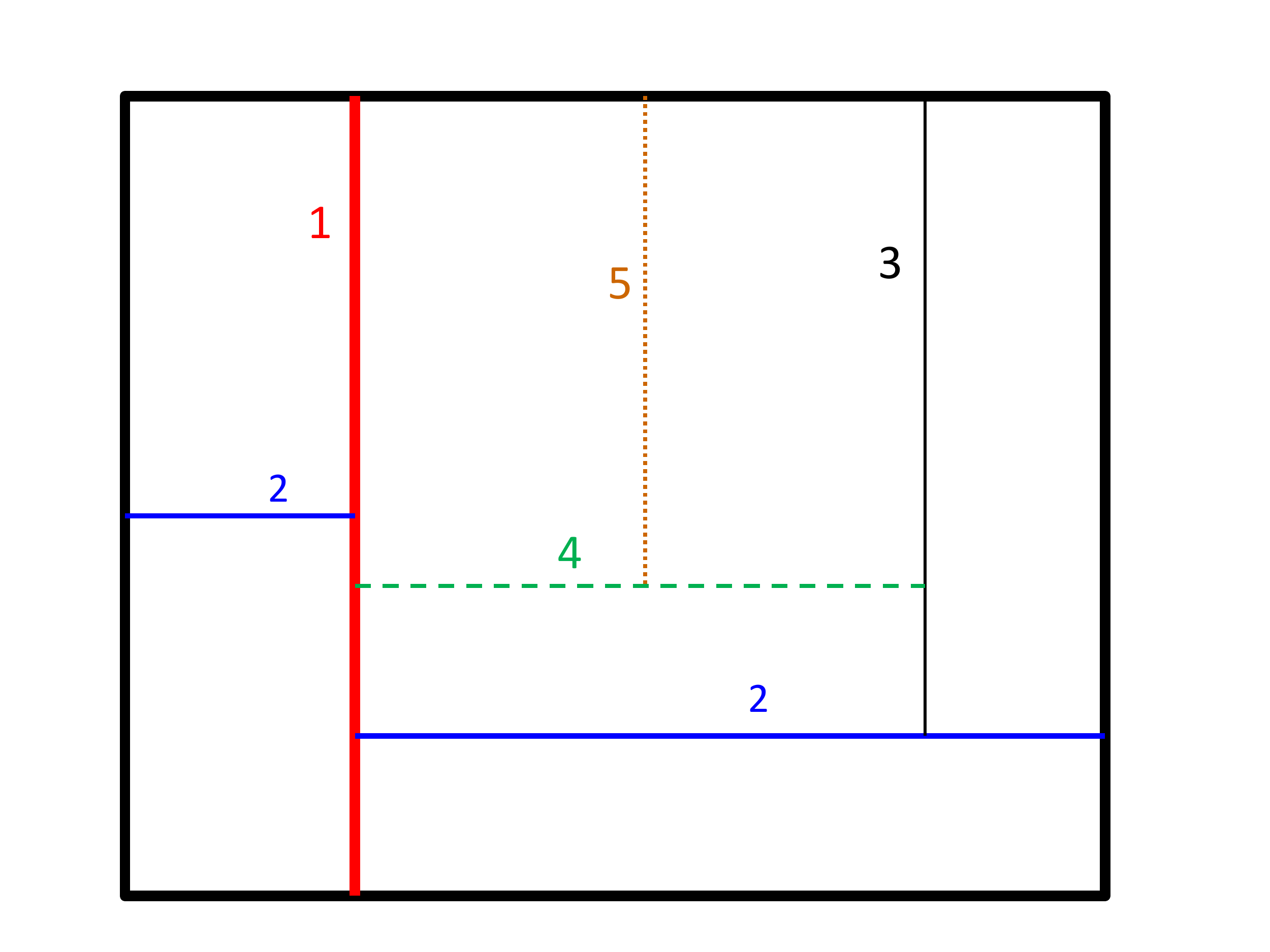}
\caption{Subsequent divisions of the network lead to the levels of a hierarchy. For the sake of clarity, here the same idea is shown for a rectangle. Boundaries of areas resulted from  subsequent partitions are marked with numbers from 1 to 5, and by color (online): red, blue, black, dotted green, dotted yellow. Boundaries made earlier are supposed to be thicker.}
\label{nested}
\end{center}
\end{figure}

During one time step, each fragment of the network which can be divided is divided. In other words, subsequent partitions are performed on each fragment of the network simultaneously. In Fig. \ref{nested}, we show the same idea, pictured for clarity on a rectangle. There, first partition is marked by a vertical line 1. Each of two parts of the rectangle are divided at step 2 - these are two horizontal blue lines marked by 2. One part of the rectangle (upper right) is divided into two by the black vertical line, etc. The final partition shown in the picture is obtained in five time steps. At each time step, the resulting parts are nested in the parent rectangle; that is why we can speak about a hierarchy of partitions.\\

\section{Numerical results}

\begin{figure}[!hptb]
\begin{center}
\includegraphics[width=.99\columnwidth, angle=0]{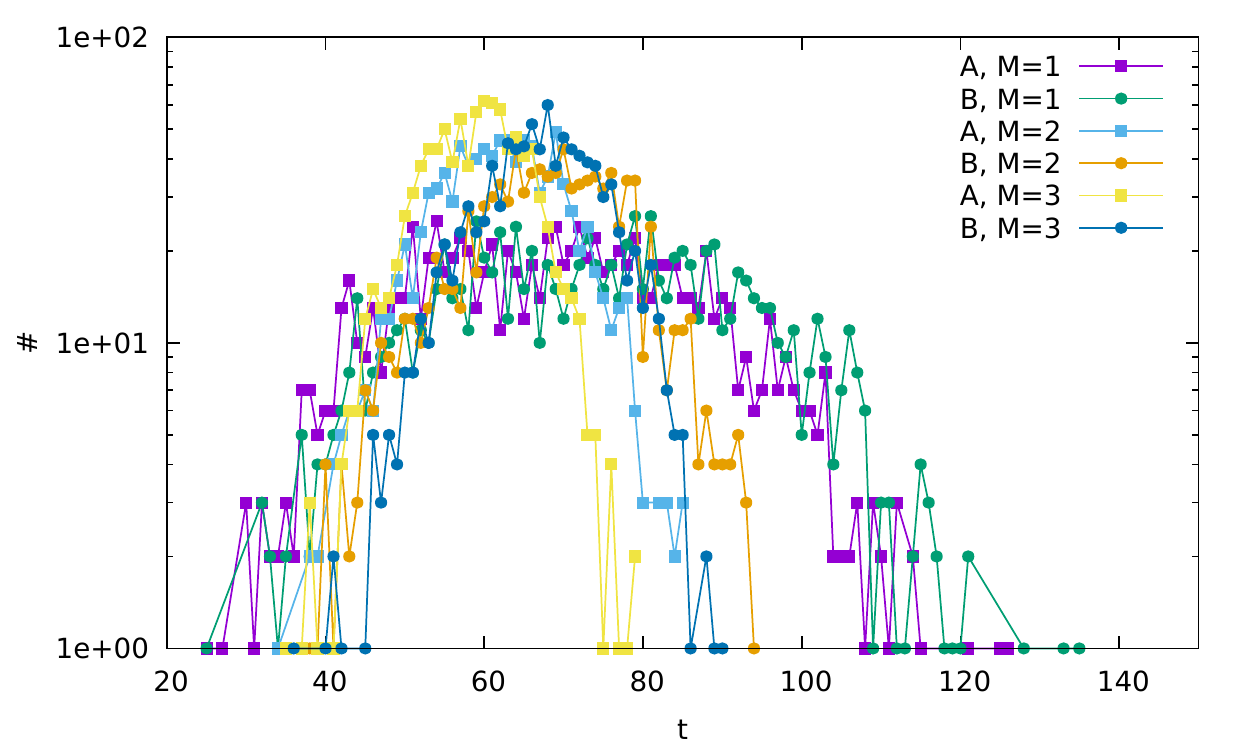}
\caption{The histogram $\sharp (t)$ of the time length $t$ of the division process for the variants A and B, and for different values of the parameter $M$.}
\label{times}
\end{center}
\end{figure}

In Fig. \ref{times}, we show the histogram $\sharp (t)$ of the duration time of the separation, measured as the number of the partitions. There is no marked difference between particular cases (different variants A and B, different values of $M$), except the upper limits of the distributions, on the right side of the histograms. There we see that basically, the larger value of $M$, the shorter time (i.e. smaller number of divisions). This result is a consequence of the fact, that for a more dense network (larger $M$), the fragment of a diameter three - which cannot be divided anymore, has more nodes, and therefore the time of division to reach such a fragment is shorter. In other words, the division of a denser network stops earlier. \\

On the other hand, the times obtained in the variant A are usually shorter than those in the variant B. This means, that cutting the links which are formed earlier is more efficient. This result is akin to the fact, that the scale-free networks are more sensitive to the removal of nodes with larger degree \cite{xxx,yyy}. By means of the algorithm of growth, these nodes are more aged, i.e. they are added earlier. \\

\begin{figure}[!hptb]
\begin{center}
\includegraphics[width=.99\columnwidth, angle=0]{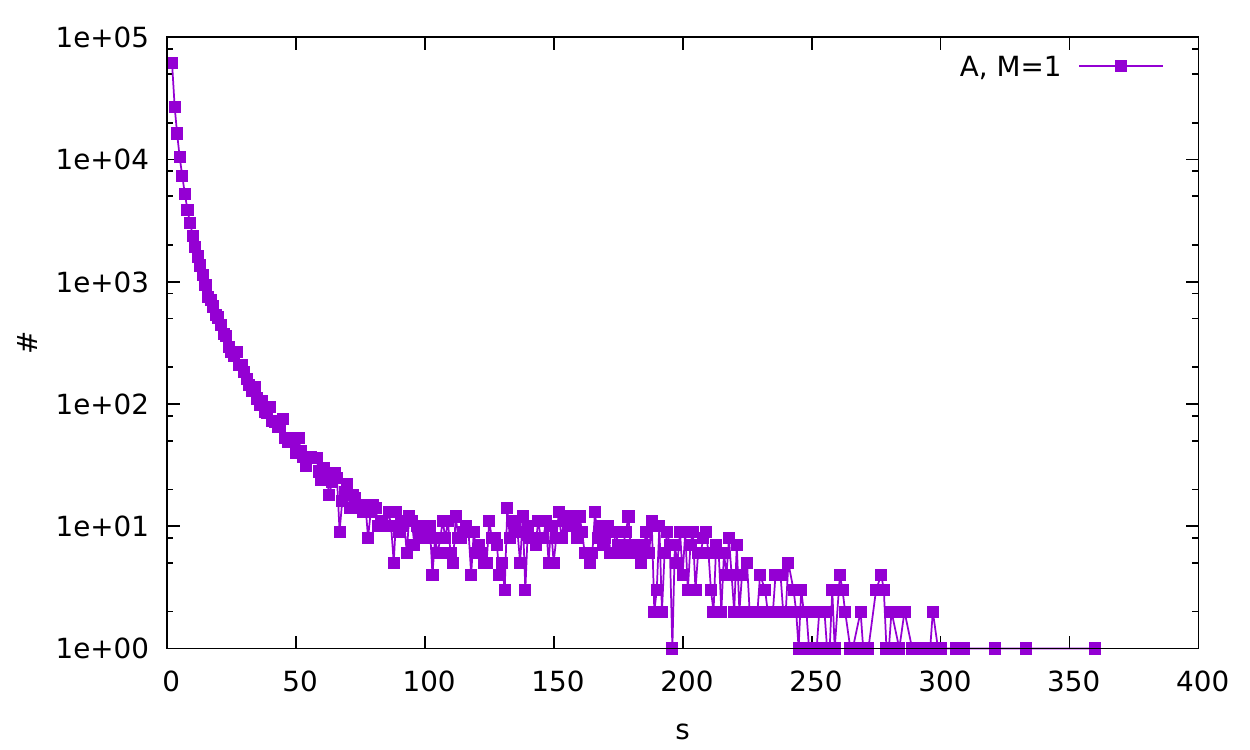}
\includegraphics[width=.99\columnwidth, angle=0]{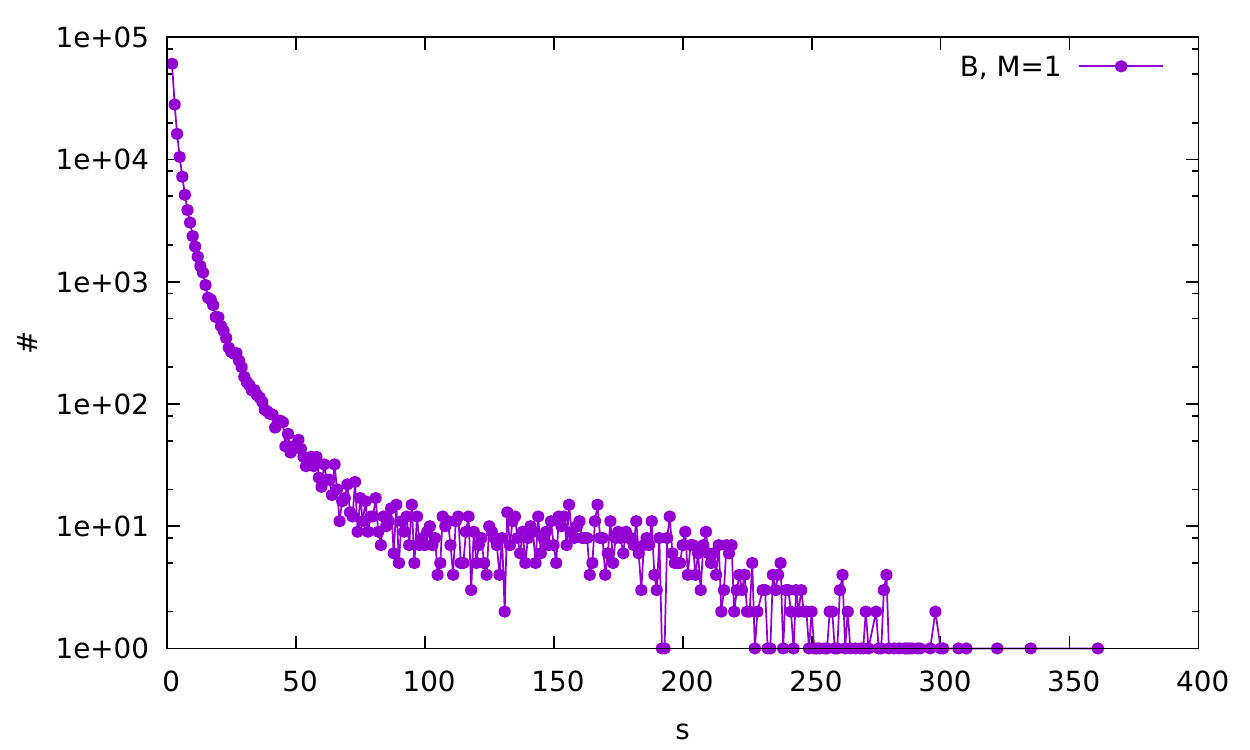}
\caption{The histogram $\sharp (s)$ of the fragment size $s$ distribution for variant A (upper plot) and B (bottom), for $M$ = 1.}
\label{si1}
\end{center}
\end{figure}

\begin{figure}[!hptb]
\begin{center}
\includegraphics[width=.99\columnwidth, angle=0]{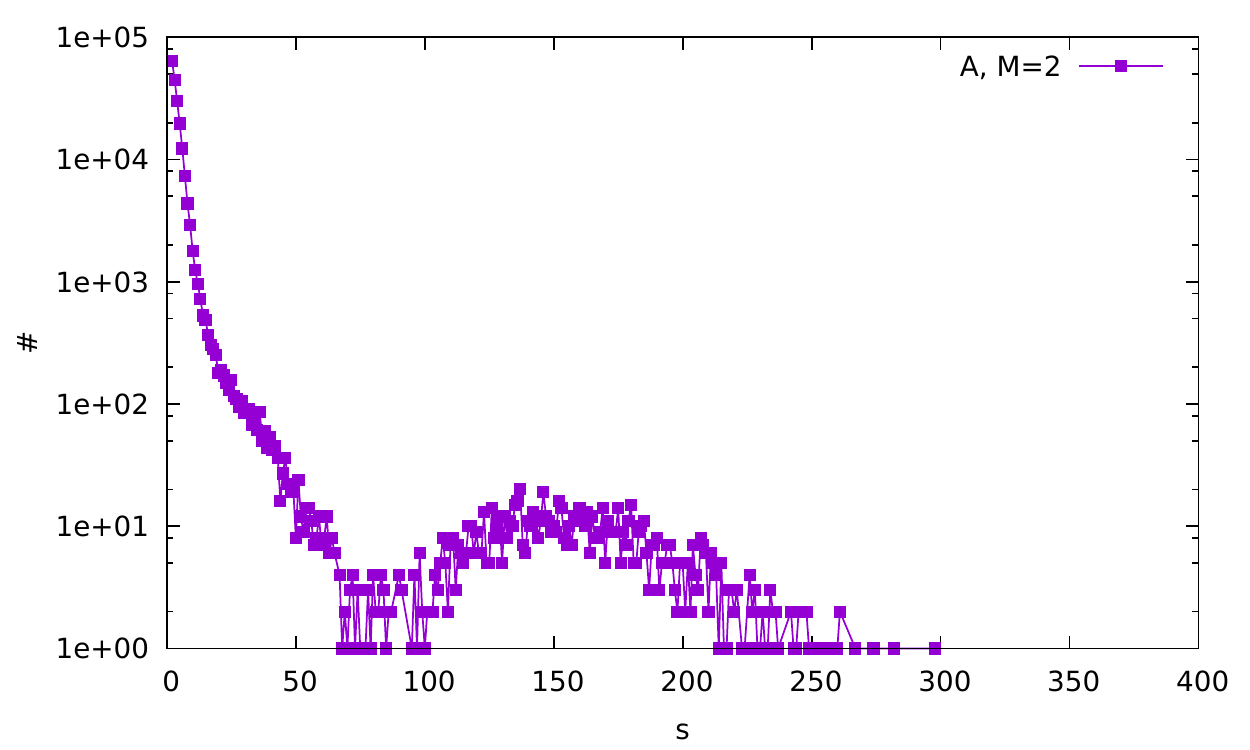}
\includegraphics[width=.99\columnwidth, angle=0]{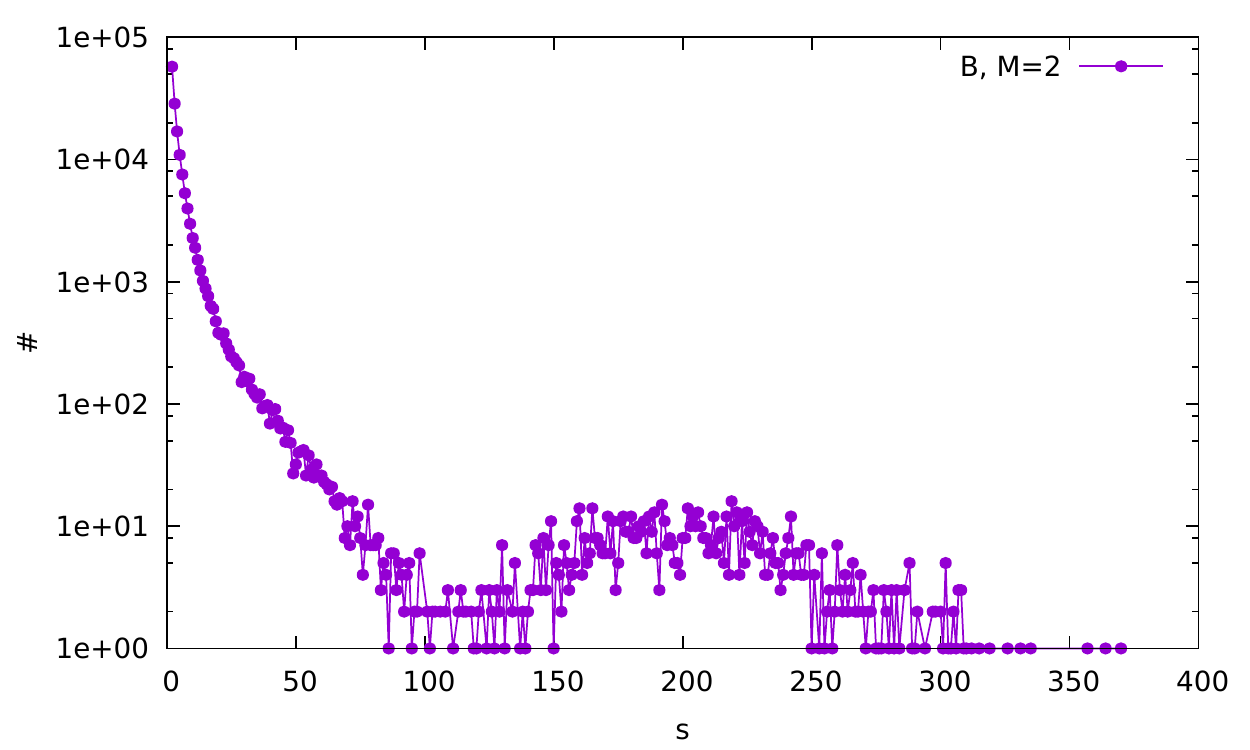}
\caption{The histogram $\sharp (s)$ of the fragment size $s$ distribution for variant A (upper plot) and B (bottom), for $M$ = 2.}
\label{si2}
\end{center}
\end{figure}

\begin{figure}[!hptb]
\begin{center}
\includegraphics[width=.99\columnwidth, angle=0]{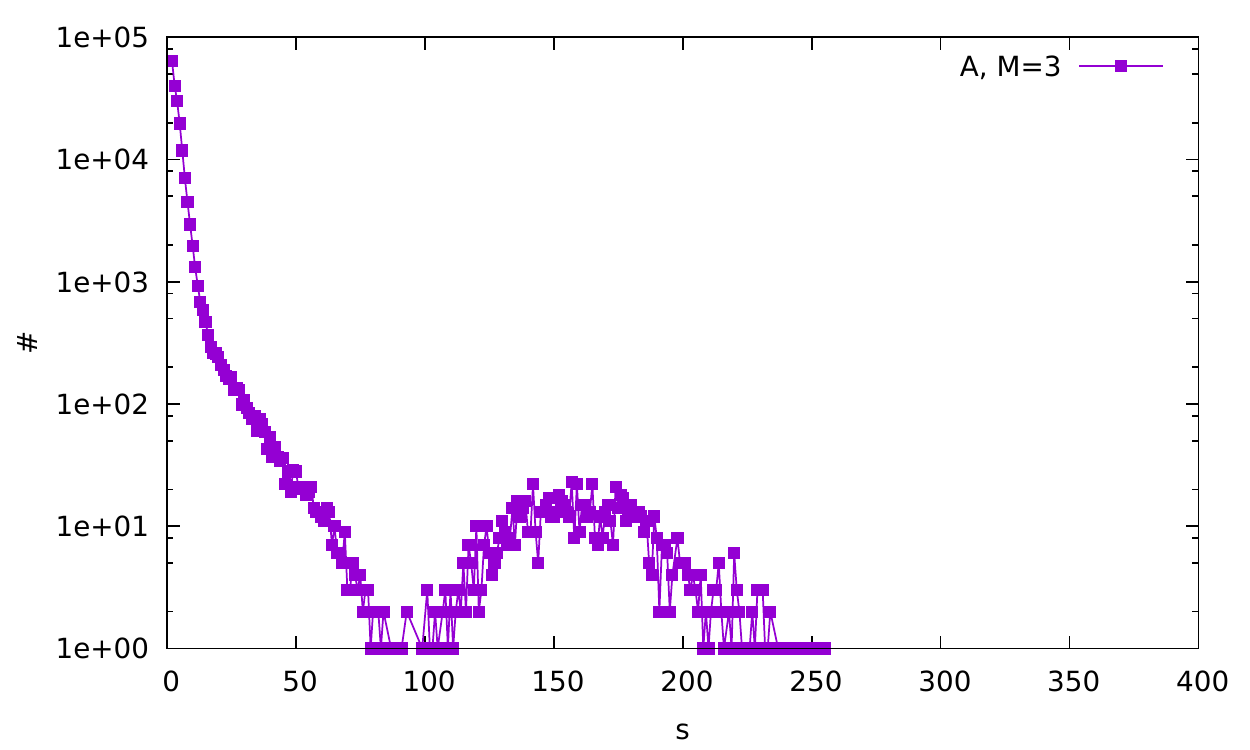}
\includegraphics[width=.99\columnwidth, angle=0]{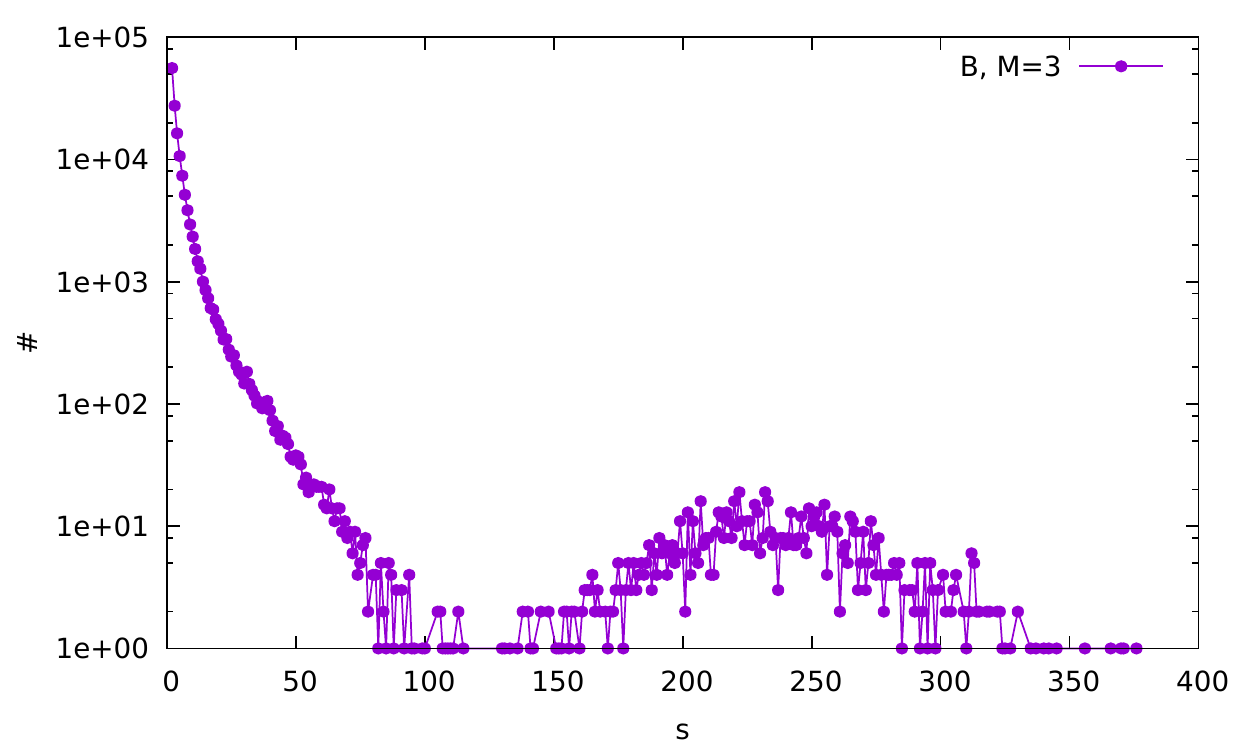}
\caption{The histogram $\sharp (s)$ of the fragment size $s$ distribution for variant A (upper plot) and B (bottom), for $M$ = 3.}
\label{si3}
\end{center}
\end{figure}

The calculations have been performed for $N$ = 1000. For both variants A and B and $M$ = 1, 2 and 3, the size distribution of the fragments of the network for the obtained final partition are shown in Figs. (\ref{si1}, \ref{si2}, \ref{si3}). It can be seen that in all cases, the distribution consists of two parts. The maximum on the right side, visible in the semilogarithmic scale, shows the largest component which contains the node with the largest degree in the whole network. The volume of this peak is therefore equal to the number of simulated networks, which is $K=10^3$. We expect that the largest fragment contains at least the hub plus its nearest neighbours. We shall estimate analytically the size of this component in the next section.  


\begin{figure}[!hptb]
\begin{center}
\includegraphics[width=.99\columnwidth, angle=0]{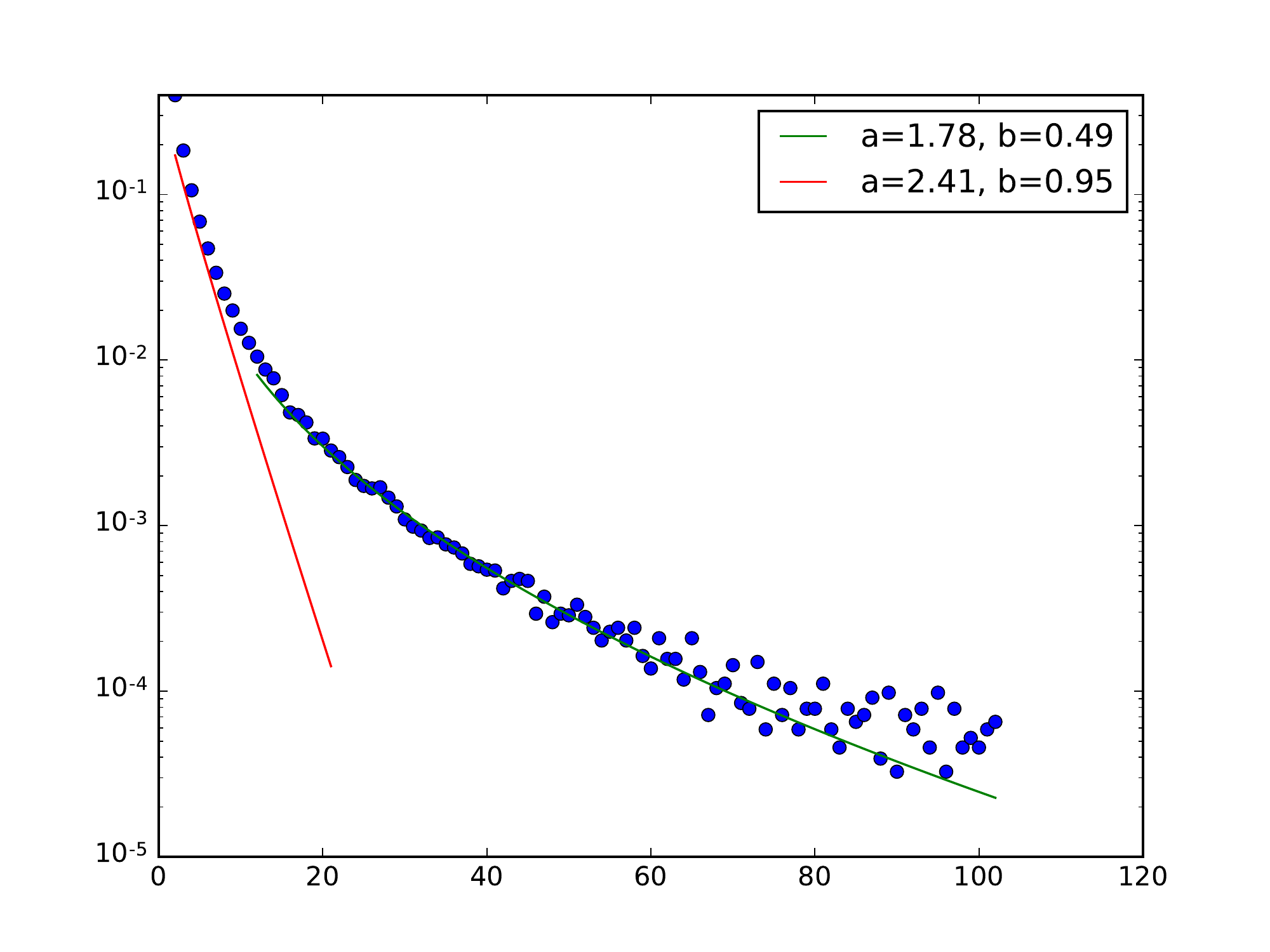}
\caption{An exemplary fit of the numerical results by the Weibull distribution for M=1, variant B.}
\label{fit}
\end{center}
\end{figure}

On the left side of the Figs. (\ref{si1}, \ref{si2}, \ref{si3}) we see the size distribution of the remaining fragments of the network. We can analyze this part independently on the maximum on the right side of the plots. Trying to fit these left parts of the plots with the two-parameter Weibull distribution \cite{wei}

\begin{equation}
\rho(s)=ab(as)^{b-1}e^{-(as)^b}
\end{equation}
we observe two ranges of $s$ where different values of the parameter $b$ are obtained. For each out of the six ($M$ = 1, 2 and 3, variants A and B) distributions, $b$ is about 0.95 for low $s$, and about 0.4-0.5 for higher $s$. The ranges of size $s$ where the different values of $b$ fit can be seen in Figs. (\ref{si1}, \ref{si2}, \ref{si3}). The parameter $a$ varies between 0.4 and 2.8. An exemplary fit is shown in Fig. \ref{fit}.

\section{Analytical estimation of mean size of giant subtree}

It is reasonable to  assume that right parts of  distributions in
Figs. \ref{si1},\ref{si2},\ref{si3} correspond to  subtrees that include the main hub of
the system. Let us estimate the {\it mean size} of  this component (we call it 'a giant subtree' here)
using a single mean-field approach. Following \cite{xxx,yyy} the
degree of a node $a$  in Albert-Barabasi network in mean-field approximation increases with
time  $t$ as
\begin{equation}
k_a(t)=M\sqrt{t/t_a} \label{JH1}
\end{equation}
  where $ t _a$ is the moment when the node $a$ was attached to the
network. This approximation is known to be mediocre for extremal values of node degree, as for a hub \cite{zb}, so our approach is not rigid.

Let assume for the simplicity that that $M=1$ thus the
network is a loopless tree. The idea is to count the hub, its nearest neighbors and their neighbors with the condition that the latter have no other neighbors; they remain as leaves. In other words, the giant subtree contains the hub, all its children, and those grandchildren that have not got their own offspring. This means that the giant component includes only the grandchildren that are not too mature. Let $\Omega_h$  stands for a  $h-$ sphere
of the hub, i.e it is  set of all nodes that are at a distant
$h=0,1,2\ldots$  from the hub. Following the assumed splitting  algorithm the giant subtree consists of   spheres
$\Omega_0$ (the hub),   $\Omega_1$ (its children),    and a part of nodes from  $\Omega_2$ (grandchildren)
that are not connected to any node in  the sphere $\Omega_3$.  Let
the total number of nodes in the network is  $t+1$, {\it i.e.} the age of
the network is $t$ (hub has been born at  $t=0$). Since the hub's  degree is  $h_{hub}(t) =
\sqrt{t}$  thus at  moments $t_i=i^2$  where  $i=1,2,3,\ldots \le
\sqrt{t}$ nodes $i$ in $\Omega_1$  emerge. Note that this reasoning {\it underestimates} the size of the giant subtree. In fact, it assumes that the hub is not increasing at all at time moments $t=2,3$, although the evolutionary algorithm of the Albert-Barabasi network works at these times as well. Following  (\ref{JH1})
degrees of these nodes increase as
\begin{equation}
k_i(t)=\sqrt{t}/i \label{JH2}
\end{equation}
Thus  at moments $t=t_{i,j}= i^2 j^2$ , $j=2,3,4,  \ldots ,\le \sqrt
t)/i $  nodes $(i,j)$ in   the sphere $\Omega_2$ emerge that are directly
connected to the node $i$ from the sphere $\Omega_1$  (let us note that
$t_{i,1}$ corresponds to the emergence of the node $i$) .
Following  (\ref{JH1}) degrees of these nodes  increase as
\begin{equation}
k_{i,j}(t)=\sqrt{t}/(ij) \label{JH3}
\end{equation}

  It means that   at the moment $t_{i,j,2}=4i^2j^2 $ the degree of the
node $(i,j)$ from $\Omega_2$ equals to  $k_{i,j}=2$, i.e.
there is already one node $(i,j,2)$ emerged in $\Omega_3$ that is  directly connected to to the node  $(i,j)$. Thus if $t\ge
4i^2j^2$ then the node $(i,j)$ is not   in the same subtree as the
hub. It means that  among all nodes  $(i,j)$ only nodes described by
pairs of  labels such that     $i=1,2,3,\ldots \le \sqrt{t}$ and  $
\sqrt{t}/(2i) \le  j  \le \sqrt t)/i$ and $j\ge 2$ are in the same
subtree as the hub. When $t\gg 1$ the number of such nodes from the
second sphere can be approximately calculated as the integral
  \begin{equation}
S^{hub}_2\approx \int_1^{\sqrt t} di \int_{\sqrt t/(2i)}^{\sqrt t/i}
dj=\sqrt t \ln (t) /4   \label{JH4}
\end{equation}
The above formula  does not take into account  the condition   $j\ge
2$  thus the number of nodes is overestimated.
  For large networks when $t \gg 1$  we can estimate the size  of the
subnetwork with the  hub as
  \begin{equation}
S^{hub}_{total}(t)\approx 1+\sqrt t+ \sqrt t \ln (t) /4 \label{JH4}
\end{equation}
  For $t=1000$ we get $S^{hub}_{total}\approx 1+31.6+ 54.6=87.2$
Unfortunately the  above value is only about  the half  of the maximum
observed in Fig. 3 for $S_{numer}^{hub} \approx 150 $.  \\

This discrepancy is supposed to follow from the approximation adopted here for the time dependence of the hub degree (Eq. \ref{JH1}). The numerical values  of $S^{hub}_{total} (t)$ for $t=500, 1000$ and $2000$ are found to be about $110, 160$ and $220$, respectively, which is proportional to the square root of $t$. In this sense numerical values of $S^{hub}_{total}(t)$ scales with the network size $t$  similarly as the formula (\ref{JH4}). Similar proportionality is obtained for $M=2$, as shown in the Supplementary Material.

\section{Application to the network of blogs}

In the virtual world, a leader can find an equivalent to a leading source of information; popularity of both can be measured by the node degree in a network \cite{wfst,ball}. Here
we use this analogy to apply our algorithm to the network of political weblogs in U.S., carried on within two months before the presidential election in 2004 \cite{adamic}. There, nodes are blogs and a link means that one blog refers to another. The blogs are tagged as democratic or republican. Our intention is not as to check if they are politically oriented, because clearly they are, but rather to verify if our algorithm based solely on the network topology can reproduce this polarization. \\

\begin{figure}[!hptb]
\begin{center}
\includegraphics[width=.99\columnwidth, angle=0]{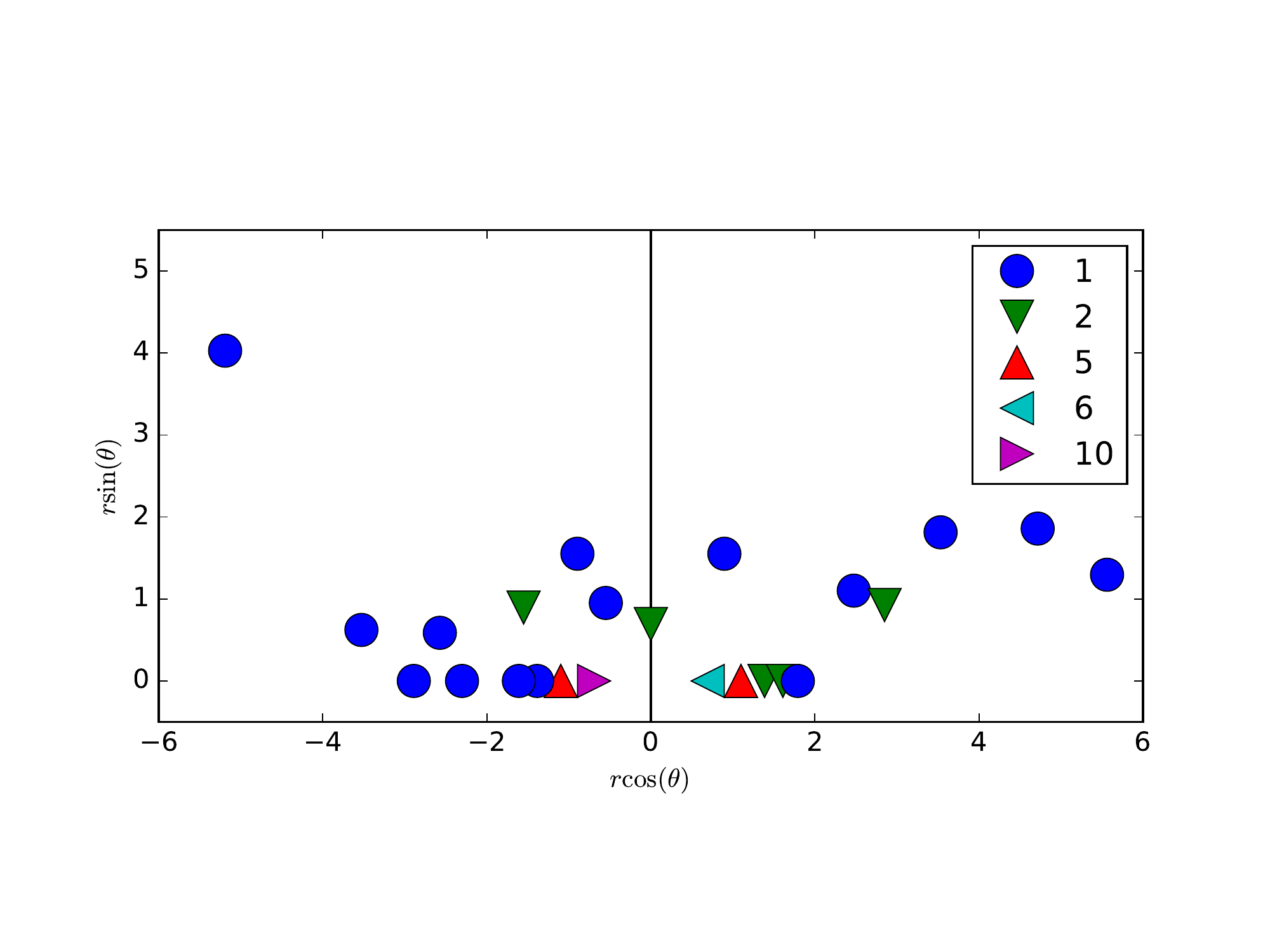}
\includegraphics[width=.99\columnwidth, angle=0]{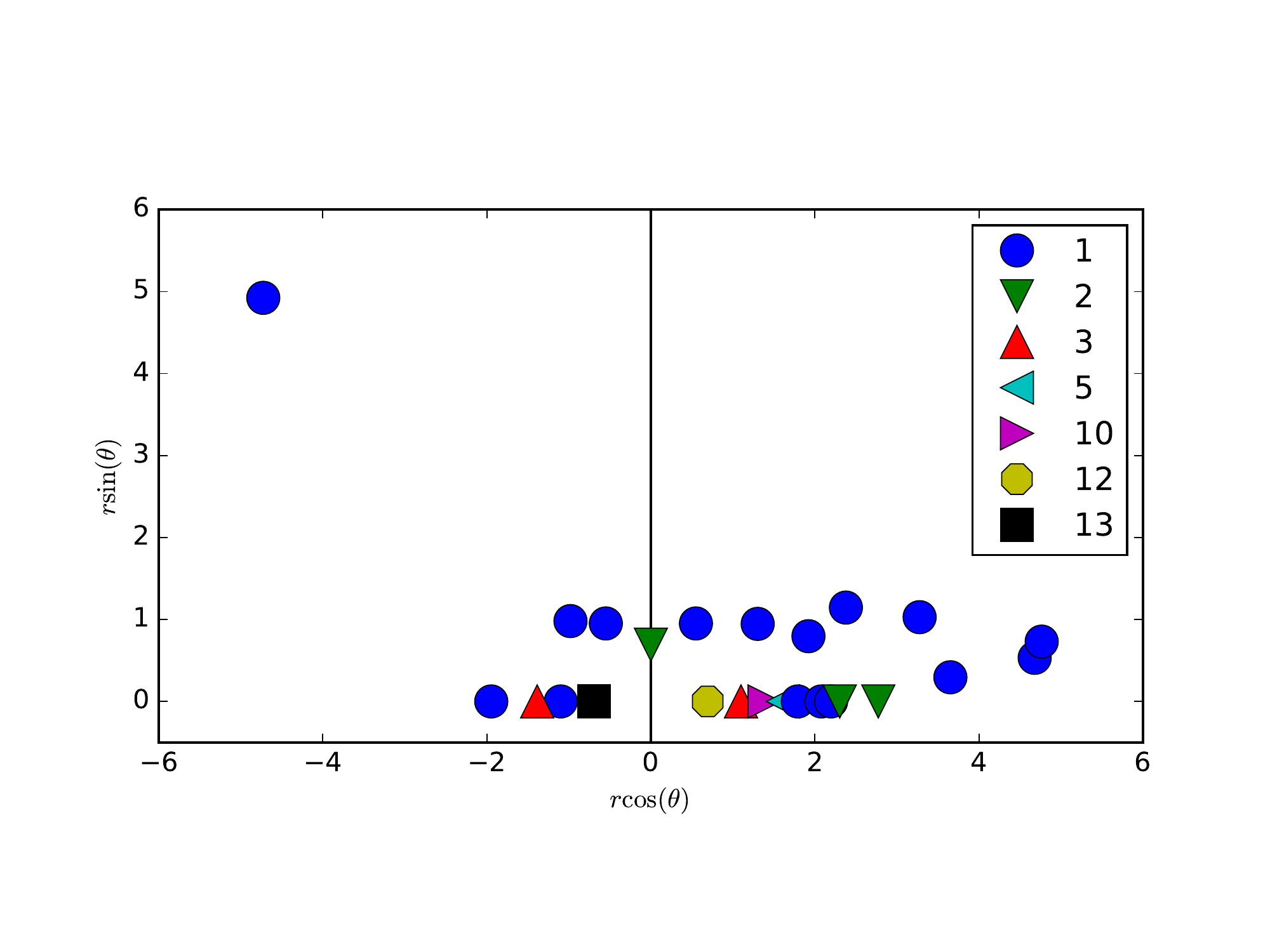}
\caption{The fragments obtained by the partition of the network of political blogs \cite{adamic}. The position of a fragment is related to its size and its content: $r=\log(N_R+N_D)$, and $\Theta= \pi N_D/(N_R+N_D)$, where $N_D$ ($N_R$) is the number of blogs tagged as democratic (republican). The same data on the blogs \cite{adamic} are used twice, according to the two versions of the algorithm: stronger (weaker) links are cut at first for the upper (the lower) picture. For some positions in the plane, we find more than one fragment, hence different symbols are used for the number of fragments at a given position (see  legends on both plots). In both figures, the blue dot on the upper left corner contains the largest hub.}
\label{blogi}
\end{center}
\end{figure}

For our purposes, links are symmetrized. Yet, if two blogs refer to each other, the weight of their mutual link is 2 (a bidirectional reference); otherwise it is either 1 (an unidirectional reference) or zero (no reference). These weights allow to perform the simulation in two ways, cutting stronger or weaker links at first, as we previously did with younger or older links. In Fig.(\ref{blogi}) the distribution of sizes and political orientations of the obtained fragments are depicted for both versions of the algorithm. \\

As we see in Fig. \ref{blogi}, the majority of fragments is fully polarized, i.e. either the number of blogs tagged as democratic $N_D$  or republican  $N_R$ in a given fragment is equal to zero. This corresponds to the value of the parameter $\Theta= \pi N_D/(N_R+N_D)$  either zero or $\pi$. Even the largest fragments are clearly polarized; their partition shows a clear majority of democratic blogs. Concluding, in all but some fragments, this or that political orientation prevails. It is also worthwhile to note that the nodes which play the role of leaders (hubs) at first step of separation belong to different parties in both variants of the algorithm. Also, the fragments which contain these hubs are the largest ones. In the largest fragment which contains the largest hub, on the upper picture: $N_D=684$, $N_R=236$, and on the lower picture: $N_D=564$, $N_R=150$. In the second largest fragment, which contains the second largest hub, in the upper picture: $N_D=6$, $N_R=118$, and in the lower picture: $N_D=22$, $N_R=280$. We note that in the absence of polarization, the most probable partition would be  $N_D \approx N_R$.\\

\section{Discussion}

Our simulations are performed for two versions of the algorithm, where old or new links are cut at first. Yet, there are no qualitative differences between the results. We deduce that any intermediate version of the algorithm should produce similar results; in other words, the order of cutting links is not that crucial. The same conclusion applies also to the order of cutting, uni- or bidirectional, in the network of blogs.\\

The observed distribution of sizes of splitted componets consists of two parts. Smaller components follow  the Weibull distribution, known also as the generalized Mott distribution,  that has been applied to describe the size distribution of the fragments of explosive warheads \cite{elep}.
The difference of the parameters $b$ of the Weibull distributions we have found in different ranges of the fragment size $s$ indicates, that small fragments are produced according to a different rule, than larger ones. We deduce that these small fragments come from  the surface of the largest fragment containing the first leader. Their particular distribution is due to the condition that the nearest neighbours of the leader remain attached to him. 
Largest fragments follow from  the structure of the closest neigbourhood of the main network hub or the group primary group leader. Our numerical  simulations  as well as  mean-field theory show that  size of the largest fragment scales approximately as the square root of the initial network size. The results of the application of the algorithm to the network of politically polarized blogs confirm that the obtained network partition properly reflects the conflict, encoded in the network structure. \\

We note that if our results are relevant for social reality, they are related to possible partitions as well as to those which actually occurred. A leader who analyses a partition which might happened prefers a scenario, when the part of network ruled by his rival is splintered into parts at the beginning of the process. Then, a prudent leader tries in advance to set his opposition at variance; two weak rivals are better than a strong one. This strategy is known throughout the whole human history \cite{kers,bs,gjc}. Our considerations can be seen as a mathematical illustration of importance of this strategy.\\

\vspace{1cm}
\noindent
{\bf Acknowledgements}\\
 The work was partially supported by the PL-Grid Infrastructure ,  as {\it RENOIR} Project by the European Union’s Horizon 2020 research and innovation programme under the Marie Sk\l odowska-Curie grant agreement No 691152  and by Ministry of Science and Higher Education (Poland), grant Nos. 34/H2020/2016,   329025/PnH /2016.   J.A.H. has been partially supported by the Russian Scientific Foundation, proposal \#14-21-00137. 
 


\begin{thebibliography}{99}

\bibitem{tatar} W. Tatarkiewicz, {\it History of Philosophy} (in Polish), PWN, Warszawa 1988.
\bibitem{heat} Ibn Khaldun, {\it The Muqaddimah: An Introduction to History}, Princeton UP, Princeton 1967. 
\bibitem{duin} J. Duindam, {\it Dynasties: A Global History of Power, 1300-1800}, Cambridge UP, Cambridge 2016, p. 296.
\bibitem{b1} S. A. Rands, G. Cowlishaw, R. A. Pettifor, J. M. Rowcliffe and R. A. Johnstone, BMC Evolutionary Biology {\bf 8}, 51 (2008).
\bibitem{b2} J. L. Harcourt, T. Z. Ang, G. Sweetman, R. A. Johnstone and A. Manica, Current Biology {\bf 19}, 248 (2009).
\bibitem{barth} F. Barth, J. of the Royal Anthropological Institute {\bf 89}, 5 (1959).
\bibitem{specu} B. B. Lichtenstein and D. A. Plowman, The Leadership Quarterly {\bf 20}, 617 (2009).
\bibitem{kers} K. Kersten, {\it The Establishment of Communist Rule in Poland, 1943-48}, Univ. of California Press, Berkeley 1991.
\bibitem{raf} R. A. Francisco, {\it Collective Action Theory and Empirical Evidence}, Springer Science+Business Media, LLC 2010, p. 99.
\bibitem{zac} W. W. Zachary, J. of Anthropological Research {\bf 33}, 452 (1977).
\bibitem{sus} C. R. Sunstein, J. of Political Philosophy {\bf 10}, 175 (2002).
\bibitem{opp} K.-D. Opp, in {\it Social Norms}, M. Hechter and K.-D. Opp (Eds.), Russell Sage Foundation, New York 2001.
\bibitem{for} S. Fortunato, Physics Reports {\bf 486}, 75 (2010).
\bibitem{T1} J. Toruniewska, K. Suchecki and J. A. Ho{\l}yst, Physica A {\bf 460}, 1 (2016).
\bibitem{L1}  K. Kacperski, J.A. Ho{\l}yst, Journal of Statistical Physics,   {\bf 84}, 169 (1996).
\bibitem{L2} J.A. Ho{\l}yst, K. Kacperski and F. Schweitzer, Annual Reviews of Computational Physics, Vol. IX, World Scientific, Singapore, 2001, p. 275;
\bibitem{L3} K. Kacperski and J.A. Ho{\l}yst, Physica A {\bf 287}, 631 (2000).
\bibitem{wfst} S. Wasserman and K. Faust, {\it Social Network Analysis: Method and Applications}, Cambridge University Press, Cambridge 1994, p. 178.
\bibitem{ball} B. Ball and M. E. J. Newman, Network Science {\bf 1}, 16 (2013).
\bibitem{jcm} J. C. Maxwell, {\it The 21 Irrefutable Laws of Leadership: Follow Them and People Will Follow You}, Thomas Nelson, Nashville 2007.
\bibitem{hb} {\it Encyclopedia of Early Christianity}, E. Ferguson (Ed.), Garland Publishing Inc., New York 1998.
\bibitem{is} F. Buhl and A. T. Welch, in {\it The Encyclopaedia of Islam. New Edition}, C. E. Bosworth, E. Van Donzel, W. P. Heinrichs and G. Lecomte (eds.),E. J. Brill, Leiden 1993.
\bibitem{sciam} A.-L. Barabasi and E. Bonabeau, Sci. Am. {\bf 288}, 50 (2003).
\bibitem{adamic} L. A. Adamic and N. Glance, {\it The political blogosphere and the 2004 US election: divided they blog}, Proc. of the WWW-2005 Workshop of the Weblogging Ecosystem (2005).
\bibitem{bk} M. E. J. Newman, {\it Networks. An Introduction}, Oxford University Press, Oxford 2010.
\bibitem{xxx} R. Albert, H. Jeong and A.-L. Barabasi, Nature {\bf 406}, 378 (2000).
\bibitem{yyy} R. Albert, A.-L. Barabasi, Rev. Mod. Phys. {\bf 74}, 47 (2002).
\bibitem{qq} S. N. Dorogovtsev and J. F. F. Mendes, Adv. Phys. {\bf 51}, 1079 (2002).
\bibitem{wei} E. A. Cohen, Jr., Mathematical Modelling {\bf 2}, 19 (1981).
\bibitem{zb} Z. Burda, private information.
\bibitem{elep} P. Elek and S. Jaramaz, 23-rd Int. Symp. on Ballistics, Tarragona, Spain, 16-20 April 2007.
\bibitem{bs} B. Simms, {\it Europe: The Struggle for Supremacy from 1453 to the Present}, Basic Books, New York 2013.
\bibitem{gjc} J. Caesar, {\it The Gallic Wars}, classic.mit.edu/Caesar/gallic.html

\end{thebibliography}
\end{document}